%% file: cameraready.tex
\pdfoutput=1

\documentclass[11pt]{article}

\usepackage{acl}

\usepackage{times}
\usepackage{latexsym}

\usepackage[T1]{fontenc}

\usepackage[utf8]{inputenc}

\usepackage{microtype}

\usepackage{inconsolata}

\usepackage{graphicx}
\usepackage{algorithm}
\usepackage{algorithmic}
\usepackage{booktabs, multirow}
\usepackage{lipsum}
\usepackage{amsmath}
\usepackage{amsfonts}
\usepackage{hyperref}

\newcommand{\model}{\texttt{SymTax}}

\title{SymTax: Symbiotic Relationship and Taxonomy Fusion for\\Effective Citation Recommendation}

\author{Karan Goyal \\
  IIIT Delhi, India \\
  \texttt{\small karang@iiitd.ac.in} \And
  Mayank Goel \\
  NSUT Delhi, India \\
  \texttt{\small mayank.co19@nsut.ac.in} \And
  Vikram Goyal \\
  IIIT Delhi, India \\
  \texttt{\small vikram@iiitd.ac.in} \And
  Mukesh Mohania \\
  IIIT Delhi, India \\
  \texttt{\small mukesh@iiitd.ac.in}
  }

\begin{document}
\maketitle
\begin{abstract}
Citing pertinent literature is pivotal to writing and reviewing a scientific document. Existing techniques mainly focus on the local context or the global context for recommending citations but fail to consider the actual human citation behaviour. We propose \texttt{SymTax}\footnote{Accepted in ACL 2024}, a three-stage recommendation architecture that considers both the local and the global context, and additionally the taxonomical representations of query-candidate tuples and the \textit{Symbiosis} prevailing amongst them. 
\texttt{SymTax} learns to embed the infused taxonomies in the hyperbolic space and uses hyperbolic separation as a latent feature to compute query-candidate similarity. We build a novel and large dataset \texttt{ArSyTa} containing $8.27$ million citation contexts and describe the creation process in detail. We conduct extensive experiments and ablation studies to demonstrate the effectiveness and design choice of each module in our framework. Also, combinatorial analysis from our experiments shed light on the choice of language models (LMs) and fusion embedding, and the inclusion of section heading as a signal. Our proposed module that captures the symbiotic relationship solely leads to performance gains of $26.66$\% and $39.25$\% in Recall@$5$ w.r.t. SOTA on \texttt{ACL-200} and \texttt{RefSeer }datasets, respectively. The complete framework yields a gain of $22.56$\% in  Recall@$5$ wrt SOTA on our proposed dataset. The code and dataset are available at \href{https://github.com/goyalkaraniit/SymTax}{https://github.com/goyalkaraniit/SymTax}. 
\end{abstract}

\section{Introduction}

\begin{figure}[t]
    \centering
    \includegraphics[width=\columnwidth]{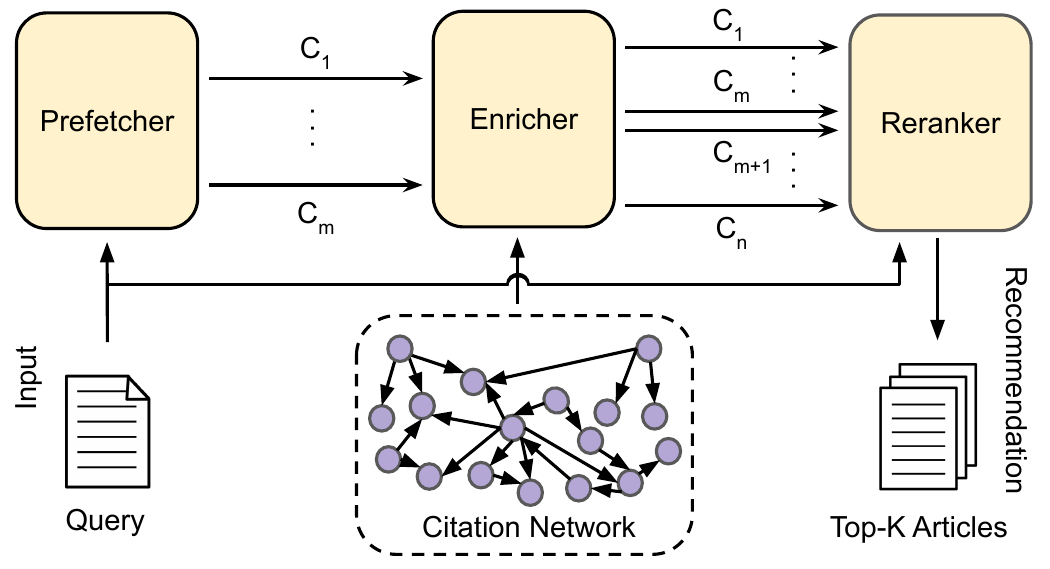}
    \caption{Proposed method consists of three essential modules. {\em Prefetcher} and {\em Reranker} takes query consisting of citation context, title, abstract and taxonomy of the citing paper as input. For each candidate paper $(C_i)$, {\em Enricher} uses knowledge from citation network and {\em Reranker} generates the final top-K recommendations.  
    }
    \label{fig:pipelineoverview}
    \vspace{-4mm}
\end{figure}

Citing has always been the backbone of scientific research. It enables trust and supports the claims made in the scientific document. The ever-growing increase in the amount of scientific literature makes it imperative to ease out the author's task of finding a list of suitable papers to follow and cite \citep{johnson2018stm,bornmann2021growth,nane2023covid}. Citation recommendation is such a process that helps researchers to be aware of the relevant research in respective domains. There are two different approaches to recommend citations: \textit{local} \citep{dai2019attentive,ebesu2017neural,huang2012recommending,he2010context}, and \textit{global} \citep{xie2021graph,10.1016/j.eswa.2021.114888,bhagavatula-etal-2018-content,guo2017exploiting}. Local citation recommendation is the task of finding and recommending the most relevant prior work, mainly corresponding to a specific text passage (also known as citation context), making it context-aware. On the other hand, global citation recommendation recommends a list of suitable prior art for the entire document, mainly given the title and abstract or the whole document. In this paper, we solve the task of local citation recommendation, which is more fine-grained and provides a solution to the actual challenge the author faces. For example, consider the below citation excerpt:\footnote{Excerpt is borrowed from \textit{Towards Consistency in Adversarial Classification} of \citep{NEURIPS2022_38d6af46}. The cited article is \textit{ An analysis of adversarial attacks and defenses on autonomous driving models} of \citep{deng2020analysis}.}

\textit{This can have extreme consequences in real-life scenarios such as autonomous cars} \texttt{CitX}.

Examining the above context in isolation makes it challenging to predict the specific article cited at \texttt{CitX}. However, leveraging global information such as title, abstract, and taxonomy narrows down the search space while at the same time utilizing symbiotic relationship provides the model with an enriched pool of the most suitable candidates. Unlike \texttt{ACL-200} and \texttt{RefSeer} datasets with curated contexts of fixed size, we curate richer contexts by incorporating complete information of adjoining sentences with respect to the citation sentence.
To summarise, we make the following contributions:
\begin{itemize}
    \item \textit{Dataset}: We have constructed a dataset \texttt{ArSyTa} comprising $8.27$ million comprehensive citation contexts across diverse domains, featuring richer density and relevant features, including taxonomy concepts, to facilitate the task of citation recommendation.    
    \item \textit{Conceptual}: We explore the concept of \textit{Symbiosis} from Biology and draw its analogy with human citation behaviour in the scientific research ecosystem and select a better pool of candidates.
    \item \textit{Methodological}: We propose a novel taxonomy fused reranker that subsequently learns projections of fused taxonomies in hyperbolic space and utilises hyperbolic separation as a latent feature.   
    \item \textit{Empirical}:  We perform extensive experiments, ablations, and analysis on five datasets and six metrics, demonstrating \texttt{SymTax} consistently outperforms SOTA by huge margins.
\end{itemize}

\section{Related Work}
Local citation recommendation has drawn comparatively less interest than its global counterpart until recently. \citet{he2010context} introduced the task of local citation recommendation by using tf-idf based vector similarity between context and cited articles. \citet{livne2014citesight} extracted hand-crafted features from the citation excerpt and the remaining document text, and developed a system to recommend citations while the document is being drafted. The neural probabilistic model of \citep{huang2015neural} determines the citation probability for a given context by jointly embedding context and all articles in a shared embedding space. \citet{ebesu2017neural} proposed neural citation network based on encoder-decoder architecture. The encoder obtains a robust representation of citation context and further augments it via author networks and attention mechanism, which the decoder uses to generate the title of the cited paper. \citet{dai2019attentive} utilised stacked denoising autoencoders for representing cited articles, bidirectional LSTMs for citation context representation and attention principle over citation context to enhance the learning ability of their framework.

\citet{jeong2020context} proposed a BERT-GCN model which uses BERT \citep{kenton2019bert} to obtain embeddings for context sentences, and Graph Convolutional Network \citep{kipf2017semisupervised} to derive embeddings from citation graph nodes. The two embeddings are then concatenated and passed through a feedforward neural network to obtain relevance between them. However, due to the high cost of computing GCN, as mentioned in \citet{gu2022local}, BERT-GCN model was evaluated on tiny datasets containing merely a few thousand citation contexts. It highlights the limitation of scaling such GNN models for recommending citations on large datasets.

\citet{medic2020improved} suggested the use of global information of articles along with citation context to recommend citations. It computes semantic matching score between citation context and cited article text, and bibliographic score from the article's popularity in the community to generate a final recommendation score. \citet{ostendorff-etal-2022-neighborhood} perform neighbourhood contrastive learning over the full citation graph to yield citation embeddings and then uses k-nearest neighbourhood based indexing to retrieve the top recommendations. The most recent work in local citation recommendation by \citet{gu2022local} proposed a two-stage recommendation architecture comprising a fast prefetching module and a slow reranking module. We build upon work of \citet{gu2022local} by borrowing their prefetching module and designing a novel reranking module and another novel module named \texttt{Enricher} that fits between Prefetcher and Reranker. We name our model as \texttt{SymTax} (\underline{Sym}biotic Relationship and \underline{Tax}onomy Fusion).

\section{Proposed Dataset}
\label{section:proposed_data}
\paragraph{Motivation.}
Citation recommendation algorithms depend on the availability of the labelled data for training. However, curating such a dataset is challenging as full pdf papers must be parsed to extract citation excerpts and map the respective cited articles. Further, the constraint that cited articles should be present in the corpus eliminates a large proportion of it, thus reducing the dataset size considerably. e.g. \texttt{FullTextPeerRead} \citep{jeong2020context} and \texttt{ACL-200} \citep{medic2020improved} datasets contain only a few thousand papers and contexts. \texttt{RefSeer} \citep{medic2020improved} contains $0.6$ million papers published till $2014$ and hence is not up-to-date. \citet{gu2022local} released a large and recent arXiv-based dataset (we refer to it as \texttt{arXiv(HAtten)}) by following the same strategy adopted by \texttt{ACL-200} and \texttt{FullTextPeerRead} for extracting contexts. They consider $200$ characters around the citation marker as the citation context. The above mentioned datasets have limited features, which may restrict the design of new algorithms for local citation recommendation. Thus, we propose a novel dataset \texttt{ArSyTa}\footnote{ArSyTa: \underline{Ar}xiv \underline{Sy}mbiotic Relationship \underline{Ta}xonomy Fusion} which is latest, largest and contains rich citation contexts with additional features.

\input{Tables/datastats}

\paragraph{\textbf{Dataset Creation.}}
We selected $475,170$ papers belonging to Computer Science (CS) categories from over $1.7$ million scholarly papers spanning STEM disciplines available on arXiv. The papers are selected from April $2007$-January $2023$ publication dates to ensure current relevance. arXiv contains an extensive collection of scientific papers that offer innate diversity in different formatting styles, templates and written characterisation, posing a significant challenge in parsing pdfs. We comprehensively evaluate established frameworks, namely, arXiv Vanity\footnote{https://github.com/arxiv-vanity/arxiv-vanity}, CERMINE\footnote{https://github.com/CeON/CERMINE}, and GROBID\footnote{https://github.com/kermitt2/grobid\_client\_python}, for data extraction. arXiv Vanity converts pdfs to HTML format for data extraction but produces inconsistent results, thus turning extraction infeasible in this scenario. CERMINE uses JAVA binaries to generate BibTeX format from pdf but fails to extract many references, thereby not providing the required level of information. GROBID is a state-of-the-art tool that accurately and efficiently produces easy-to-parse results in XML format with a standard syntax. We conduct extensive manual testing to assess parsing efficacy and finally choose GROBID as it adeptly parses more than $99.99$\% (i.e., $474,341$) of the documents. We organise the constructed dataset into a directed graph. Nodes within the graph encapsulate a rich array of attributes, encompassing abstracts, titles, authors, submitters, publication dates, topics, categories within CS, and comments associated with each paper. Edges within graph symbolise citations, carrying citation contexts and section headings in which they appear. This provides a format that offers better visualisation and utilisation of data. 

Unlike previously available datasets, which use a $200$-character length window to extract citation context, we consider one sentence before and after the citation sentence as a complete citation context. We create a robust mapping function for efficient data retrieval. Since every citation does not contain a Digital Object Identifier, mapping citations to corresponding papers is challenging. The use of several citation formats and the grammatical errors adds a further challenge to the task. To expedite title-based searches that associate titles with unique paper IDs, we devise an approximate mapping function based on LCS (Longest Common Substring), but the sheer size of the number of papers makes it infeasible to run directly, as each query requires around 10 seconds. Finally, to identify potential matches, we employ an approximate hash function called MinHash LSH (Locality Sensitivity Hashing), which provides the top 100 candidates with a high probability for a citation existing in our raw database to be present in the candidate list. We then utilise LCS matching with a $0.9$ similarity score threshold to give a final candidate, thus reducing the time to a few microseconds. Finally, our dataset consists of $8.27$ million citation contexts whereas the largest existing dataset, RefSeer, consists of only $3.7$ million contexts. The dataset is essentially comprised of contexts and the corresponding metadata only and not the research papers, as is the case with other datasets. Even after considering a relatively lesser number of papers as a raw source, we curated significantly more citation contexts (i.e., final data), thus showing the effectiveness of our data extraction technique. This is further supported empirically by the fact that our dataset has significantly higher values of average local clustering coefficient and average degree with respect to the other datasets (as shown in Table \ref{tab:data-stats}). Each citing paper and cited paper that corresponds to a citation context respectively belongs to a CS concept in the flat-level arXiv taxonomy that contains 40 classes. 
The distribution of category classes in arXiv taxonomy for \texttt{ArSyTa} is shown in Figure \ref{fig:category_dist} (Appendix).

\paragraph{\textbf{Technical Merits.}}
\texttt{ArSyTa} offers the following merits over the existing datasets: (i) As shown in Table \ref{tab:data-stats}, \texttt{ArSyTa} is $2.2$x and $2.6$x larger than \texttt{RefSeer} and \texttt{arXiv(HAtten)}, respectively. Also, our citation context network is more dense than all other datasets, clearly showing that our dataset creation strategy is better. (ii) It is the most recent dataset that contains papers till January $2023$. (iii) It contains longer citation contexts and additional signals such as section heading and document category. (iv) \texttt{ArSyTa} is suitable for additional scientific document processing tasks that can leverage section heading as a feature or a label. (v) \texttt{ArSyTa} is more challenging than others as it contains papers from different publication venues with varied formats and styles submitted to arXiv.

\section{SymTax Model}
\begin{figure*}[ht]
    \centering
    \includegraphics[width=\textwidth]{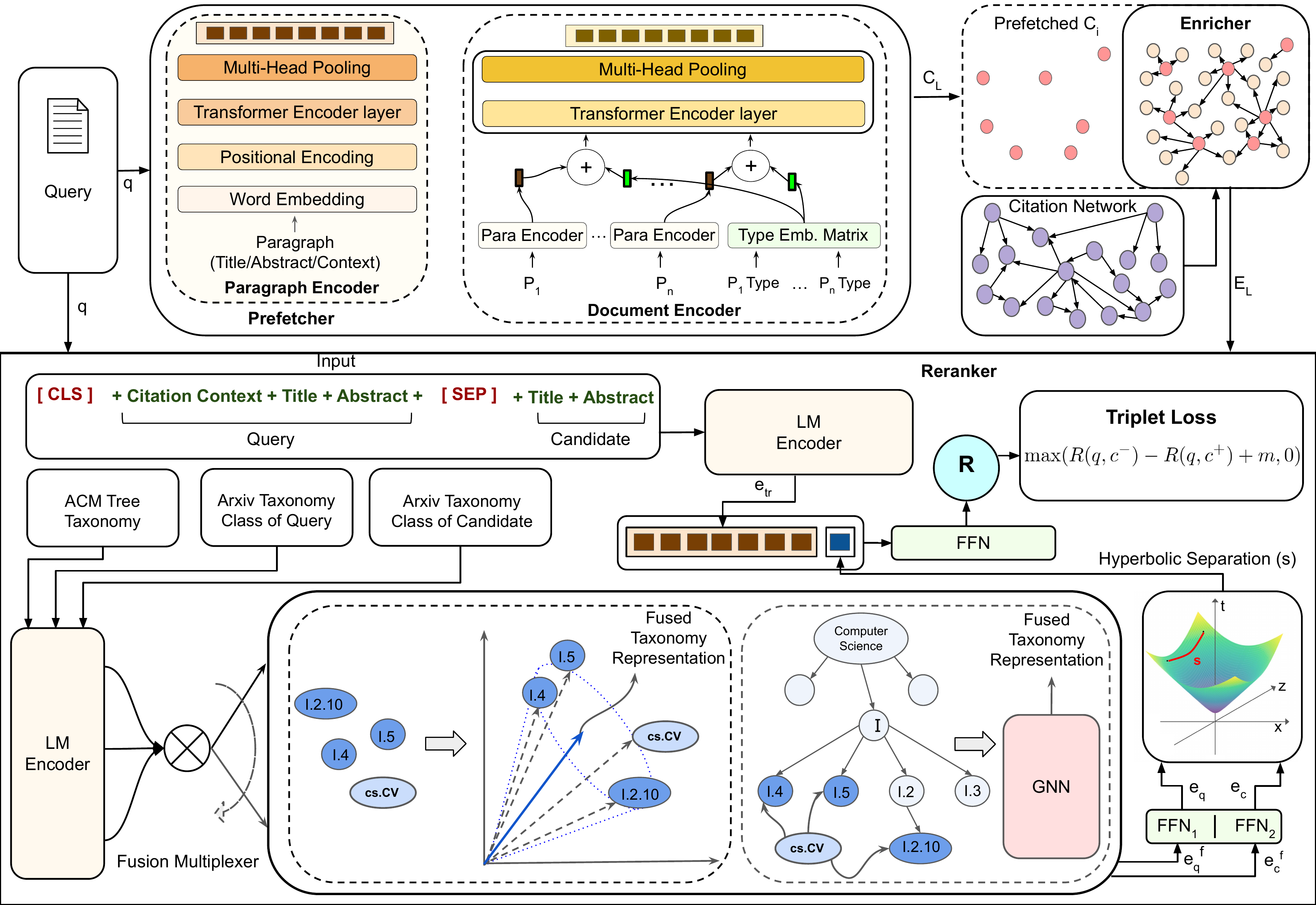}
    \caption{Architecture of \texttt{SymTax}. It consists of three essential modules -- (a) Prefetcher, (b) Enricher, and (c) Reranker. 
    The task of Enricher is to enrich the candidate list generated by Prefetcher and provide it as an input to Reranker. Reranker utilises taxonomy fusion and hyperbolic separation to yield final recommendation score (R). Mapping:- I.4: Image Processing and Computer Vision, I.5: Pattern Recognition, I.2.10: Vision and Scene Understanding, cs.CV: Computer Vision. Fusion Multiplexer enables switching between vector-based and graph-based taxonomy fusion. We have released the mapping config file along with the data.
    }
    \label{fig:model}
    \vspace{-1mm}
\end{figure*}

We discuss the detailed architecture of our proposed model -- \model, as shown in Figure \ref{fig:model}. It comprises a fast prefetching module, an enriching module and a slow and precise reranking module. We borrow an existing prefetching module from \citet{gu2022local} whereas an enriching module and a reranking module are our novel contributions in the overall recommendation technique. The subsequent subsections elaborate on the architectures of these three modules.

\subsection{Prefetcher}
The task of the prefetching module is to provide an initial set of high-ranking candidates by scoring all the papers in the database with respect to the query context. It uses cosine similarity between query embedding and document embedding to estimate the relevance between query context and the candidate document. Prefetcher comprises two submodules, namely, Paragraph Encoder and Document Encoder. Paragraph Encoder computes the embedding of a given paragraph, i.e. title, abstract or citation context, using a transformer layer followed by multi-head pooling. Document Encoder takes paragraph encodings as input along with paragraph types and passes them through a multi-head pooled transformer layer to obtain the final document embedding. We adopt the prefetching module from \citet{gu2022local} and use it as a plugin in our overall recommendation technique. For brevity, we refer readers to follow the source to understand the detailed working of the prefetcher.

\subsection{Enricher}
In general, prefetching is often followed by reranking in recommendation systems. Let $C_L = \{c_1,c_2,...,c_m\}$ be the candidate list generated by prefetcher, and $G = (V,E)$ be the given citation network $s.t.$  $c_i \in V$ $\forall$ $i \in \{1,2,...,m\}$.

The nodes in citation network represent papers and the directed edges represent citation relationship. Here, we propose a new module \texttt{Enricher} that augments the candidate list generated by the prefetcher and outputs an enriched candidate list. It generates an ego network for every article in the candidate list using the citation graph and excludes the incoming neighbours and the duplicates from the expanded network list. For every node $u \in V$, we define $N_o(u) \subset V$ as the ego network of $u$ containing only the outgoing neighbours. Let $E_L$ denotes the enriched list, then 
\begin{equation}
    \begin{split}
    E_L =& \{c_1,c_2,...,c_m,N_o(c_1),N_o(c_2),...,N_o(c_m)\} \\
    E_L =& \{c_1,c_2,...,c_m,c_{m+1},c_{m+2},...,c_n\}
    \end{split}
\end{equation}
where $\{\}$ represents a set operator. We then feed this enriched list as input to the reranker. The design notion of \texttt{Enricher} is inspired by Symbiosis, aka Symbiotic Relationship, a concept in Biology.
\vspace{-1.5mm}
\paragraph{Symbiosis.}
The idea of including cited papers of identified candidates has been pursued in the literature \citep{cohan-etal-2020-specter} but from the perspective of hard negatives. To the best of our knowledge, the concept of Enrichment has never been discussed earlier for citation recommendation to model the human citation behaviour. We identify two different types of citation behaviours that prevail in the citation ecosystem and draw a corresponding analogy with \textit{mutualism} and \textit{parasitism} that falls under the concept of \textit{Symbiosis}. \textit{Symbiosis} is a long-term relationship or interaction between two dissimilar organisms in a habitat. In our work, the habitat is the citation ecosystem, and the two dissimilar organisms are the candidate article and its neighbourhood.  We try to explain the citation phenomena through \textit{Symbiosis} wherein the candidate and its neighbourhood either play the role of \textit{mutualism} or \textit{parasitism}. In \textit{mutualism}, the query paper recommends either only the candidate paper under consideration or both the considered candidate paper and from its 1-hop outdegree neighbour network. On the other hand, in \textit{parasitism}, the neighbour organism feeds upon the candidate to get itself cited, i.e., the query paper, rather than citing the candidate article, in turn, recommends from its outgoing edge neighbours. This whole idea, in practice, is analogous to human citation behaviour. When writing a research article, researchers often gather a few highly relevant prior art and cite highly from their references. We can interpret this tendency as a slight human bias or highly as utilising the research crowd's wisdom. Owing to this, \texttt{Enricher} is only required at the inference stage. Nevertheless, it is a significantly important signal, as evident from the results in Table \ref{tab:main results} and Table \ref{tab:ablation study}.

\subsection{Reranker}
The purpose of the reranker is to rerank the candidates fetched from previous modules with higher accuracy. Therefore, the reranker is generally slower than the prefetcher. It makes a fine-grain comparison between the query and each candidate, calculates a recommendation score and returns it as the final output. Our proposed reranker considers the relevance between query and candidate at two separate levels, namely (i) text relevance and (ii) taxonomy relevance. In text relevance, we concatenate the query text and candidate text with a [SEP] token and pass it through a Language Model (LM) to obtain the joint embedding $e_{tr} \in \mathbb{R}^d$. Query text is the concatenation of citation context, title and abstract of the query article, whereas the candidate text is the concatenation of title and abstract of candidate paper. In taxonomy relevance, we perform taxonomy fusion and hyperbolic projections to calculate separation between query and candidate.

\paragraph{Taxonomy Fusion.}
The inclusion of taxonomy fusion is an important and careful design choice. Intuitively, a flat-level taxonomy (arXiv concepts) does not have a rich semantic structure in comparison to a hierarchically structured taxonomy like ACM. In a hierarchical taxonomy, we have a semantic relationship in terms of generalisation, specialisation and containment. Mapping the flat concepts into hierarchical taxonomy infuses a structure into the flat taxonomy. It also enriches the hierarchical taxonomy as we get equivalent concepts from the flat taxonomy. Each article in our proposed dataset \texttt{ArSyTa} consists of a feature category that represents the arXiv taxonomy\footnote{https://arxiv.org/category\_taxonomy} class it belongs to. Since ArSyTa contains papers from the CS domain, so we have a flat arXiv taxonomy. e.g. cs.LG and cs.CV represents Machine Learning and Computer Vision classes, respectively. We now propose the fusion of flat-level arXiv taxonomy with ACM tree taxonomy\footnote{https://tinyurl.com/22t2b43v} to obtain rich feature representations for the category classes. We mainly utilise the subject class mapping information mentioned in the arXiv taxonomy and domain knowledge to create a class taxonomy mapping from arXiv to ACM. e.g. cs.CV is mapped to ACM classes I.2.10, I.4 and I.5 (as shown in Fig. \ref{fig:model}). Also, we release the mapping config file in the data release phase. We employ two fusion strategies, namely vector-based and graph-based. In vector-based fusion, the classes are passed through LM and their conical vector is obtained by averaging out class vectors in feature space. In graph-based fusion, we first form a graph by injecting arXiv classes into the ACM tree and creating directed edges between them. We initialise node embeddings using LM and run Graph Neural Network (GNN) algorithm to learn fused representations. We consider GAT\citep{veličković2018graph} and APPNP\citep{gasteiger2018combining} as GNN algorithms and observe their performance as the same. The final representations of cs.\{\} nodes represent the fused representations learnt. Empirically, we can clearly observe that the fusion of concepts helps to attain significant performance gains (as shown in Table \ref{tab:ablation study}). 

\paragraph{Hyperbolic Separation.}
Hyperbolic spaces have recently gained momentum in deep learning due to their high capacity and tree-likeliness properties, thus making them suitable for learning better representations for hierarchical data \citep{ganea2018hyperbolic}. Motivated by the works of \citet{sawhney2022ciaug} and \citet{ganea2018hyperbolic}, and the realisation that Euclidean space cannot fully capture the complex characteristics of hierarchical data, we use hyperbolic distance as our metric to compute taxonomy relevance. Let $q$ and $c$ denote query and candidate, respectively, where $c \in E_L$. Let ${e_{q}}^{f}$, ${e_{c}}^{f} \in \mathbb{R}^D$ represents the fusion embeddings for $q$ and $c$ respectively. We pass fusion embeddings through a feed forward network $h_{\theta}(.)$ and then compute hyperbolic separation between query and candidate to project the embeddings into hyperbolic space. So, we obtain the following representations
\begin{equation}
\begin{split}
    e_{q} = h_{\theta}({e_{q}}^{f}) \in \mathbb{R}^d ;
    e_{c} = h_{\theta}({e_{c}}^{f}) \in \mathbb{R}^d
\end{split}
\end{equation}
and then compute the hyperbolic separation $s$ between $e_{q}$ and $e_{c}$ as follows
\begin{equation}
    s({e_{q}},{e_{c}}) = 2 \tan^{-1}\left(\left\|(-e_c) \oplus e_q\right\|\right)
\end{equation}
where $\oplus$ represents Möbius addition and for a pair of points $a, b \in \mathcal{B}$, is defined as,
\begin{equation*}
a \oplus b := \frac{(1 + 2\langle a, b \rangle + \lVert b \rVert^2)a + (1 - \lVert a \rVert^2)b}{1 + 2\langle a, b \rangle + \lVert a \rVert^2 \lVert b \rVert^2}
\end{equation*}
where $\langle \cdot, \cdot \rangle$, $\| \cdot \|$ are Euclidean inner product and norm.

\paragraph{Final Recommendation.}
We use this hyperbolic separation representing the taxonomy relevance as a latent feature and concatenate $(\odot)$ it with the text relevance embedding $e_{tr}$.
This step ensures that category classes with similar concepts in the taxonomy learn to embed themselves closely in the manifold. We then pass this relevance embedding through a feed forward network $(g_{\theta}(.))$ and apply the sigmoid activation function $(\sigma(.))$ to get the final relevance score, defined as,
\begin{equation}
    R = \sigma(g_{\theta}(e_{tr} \odot s)) \in (0,1)
\end{equation}
To interpret $R$ as the final recommendation score, we employ and minimise the Triplet loss, $L$:
\begin{equation}
    L = \max(R(q, c^-) - R(q, c^+) + m, 0)
\end{equation}
where $m$ is margin, and $c^+$ and $c^-$ are positive and negative candidates respectively.
We adopt a simple technique for mining triplets for a query. We choose cited paper as the positive candidate and randomly select papers from the candidate list as negative candidates.

\section{Experiments and Results}
This section illustrates the various baselines, evaluation metrics and datasets used to benchmark our proposed method followed by the performance comparison.

\paragraph{Baselines.}
We consider evaluating various available systems for comparison. \textbf{BM25} \citep{robertson2009probabilistic}: It is a prominent ranking algorithm, and we consider its several available implementations and choose Elastic Search implementation\footnote{https://github.com/kwang2049/easy-elasticsearch} as it gives the best performance with the highest speed. \textbf{SciNCL} \citep{ostendorff-etal-2022-neighborhood}: We use its official implementation available on GitHub\footnote{https://github.com/malteos/scincl}. \textbf{HAtten} \citep{gu2022local}: We use its official implementation available on GitHub\footnote{https://tinyurl.com/yckhe7d6}. {\bf NCN} \citep{ebesu2017neural} could have been a potential baseline; however, as reported by \citet{medic2020improved}, the results mentioned could not be replicated. 
\textbf{DualLCR} \citep{medic2020improved}: It is essentially a ranking method that requires a small and already existing list of candidates containing the ground truth, which turns it into an artificial setup that, in reality, does not exist. This unfair setup is also reported by \citet{gu2022local}, which is state-of-the-art in our task. Thus for a fair comparison, we could not consider it in comparing our final results.

\input{Tables/mainRes}

\paragraph{Evaluation Metrics.}
To stay consistent with the literature that uses Recall@$10$ and Mean Reciprocal Rank (MRR) as the evaluation metrics, we additionally use Normalised Discounted Cumulative Gain (NDCG@$10$) and Recall@K for different values of K to obtain more insights from the recommendation performance. Recall@K measures the percentage of cited papers appearing in top-K recommendations. MRR measures the reciprocal rank of the cited paper among the recommended candidates. NDCG takes into account the relative order of recommendations in the ranked list. The above metrics are averaged over all test queries, and higher values indicate better performance.


\paragraph{Performance Comparison.}
As evident from Table \ref{tab:main results}, our evaluation shows the superior performance of \texttt{SymTax} on all metrics across all the datasets. We consider two different variants of \texttt{SymTax} in our main results comparison (i) SpecG: with SPECTER \citep{cohan-etal-2020-specter} as LM and graph-based taxonomy fusion, and (ii) SciV: with SciBERT \citep{beltagy-etal-2019-scibert} as LM and vector-based taxonomy fusion. SPECTER and SciBERT are two state-of-the-art LMs trained on scientific text. SciV performs as the best model on \texttt{ACL-200}, \texttt{FullTextPeerRead}, \texttt{RefSeer} and \texttt{ArSyTa} on all metrics. SpecG performs best on \texttt{arXiv(HAtten)} on all metrics and results in a marginally less R@$20$ score than SciV. We observe the highest scores on \texttt{FullTextPeerRead} followed by \texttt{ACL-200}. It is due to the fact that these datasets lack diversity to a large extent. e.g. \texttt{FullTextPeerRead} is extracted from papers belonging to Artificial Intelligence field, and \texttt{ACL-200} contains papers published at ACL venues. In contrast, we observe the lowest scores on \texttt{ArSyTa} followed by \texttt{arXiv(HAtten)}. The common reason driving these performance trends is that both of these arXiv-based datasets contain articles from different publication venues with various formats, styles and domain areas, making the learning difficult and recommendation challenging. Our reasoning is further supported by the fact that \texttt{ArSyTa} is the latest dataset, and thus contains the maximum amount of diverse samples and is shown to be the toughest dataset for recommending citations. To summarise, we obtain performance gains in Recall@$5$ of 26.66\%, 23.65\%, 39.25\%, 19.74\%, 22.56\% with respect to SOTA on \texttt{ACL-200}, \texttt{FullTextPeerRead}, \texttt{RefSeer}, \texttt{arXiv(HAtten)} and \texttt{ArSyTa} respectively. The results show that NDCG is a tough metric compared to the commonly used Recall, as it accounts for the relative order of recommendations. Since the taxonomy class attribute is only available for our proposed dataset, we intentionally designed \texttt{SymTax} to be highly modular for better generalisation, as evident in Table \ref{tab:main results}.

\input{Tables/ablation}

\section{Analysis}
We conduct extensive analysis to assess further the modularity of \texttt{SymTax}, the importance of different modules, combinatorial choice of LM and taxonomy fusion, and the usage of hyperbolic space over Euclidean space. Furthermore, we analysed the effect of using section heading as an additional signal (shown in Appendix \ref{sec:appendix}).

\subsection{Ablation Study}
We perform an ablation study to highlight the importance of \textit{Symbiosis}, taxonomy fusion and hyperbolic space. We consider two variants of \texttt{SymTax}, namely SciBERT\_vector and SPECTER\_graph. For each of these two variants, we further conduct three experiments by (i) removing the \texttt{Enricher} module that works on the principle of \textit{Symbiosis}, (ii) not considering the taxonomy attribute associated with the citation context and (iii) using Euclidean space to calculate the separation score.

As evident from Table \ref{tab:ablation study}, \textit{Symbiosis} exclusion results in a drop of $21.40$\% and $24.45$\% in Recall@$5$ and NDCG respectively for SciBERT\_vector whereas for SPECTER\_graph, it leads to a drop of $17.84$\% and $20.32$\% in Recall@$5$ and NDCG respectively. Similarly, taxonomy exclusion results in a drop of 
$34.94$\% and $27.88$\% in Recall@$5$ and NDCG respectively for SciBERT\_vector whereas for SPECTER\_graph, it leads to a drop of $14.81$\% and $12.51$\% in Recall@$5$ and NDCG respectively. It is clear from Table \ref{tab:ablation study} that the use of Euclidean space instead of hyperbolic space leads to performance drop across all metrics in both variants. Exclusion of \textit{Symbiosis} impacts higher recall metrics more in comparison to excluding taxonomy fusion and hyperbolic space.

\subsection{Quantitative Analysis}

\input{Tables/quant1}

We consider two available LMs, i.e. SciBERT and SPECTER, and the two types of taxonomy fusion, i.e. graph-based and vector-based. This results in four variants, as shown in Table \ref{tab:LM and fusion choice analysis}. As evident from the results, SciBERT\_vector and SPECTER\_graph are the best-performing variants. So, the combinatorial choice of LM and taxonomy fusion plays a vital role in model performance. The above observations can be attributed to SciBERT being a LM trained on plain scientific text. In contrast, SPECTER is a LM trained with Triplet loss using $1$-hop neighbours of the positive sample from the citation graph as hard negative samples. So, SPECTER embodies graph information inside itself, whereas SciBERT does not.

\input{Tables/qual}

\subsection{Qualitative Analysis}
We assess the quality of recommendations given by different algorithms by randomly choosing an example. Though random, we choose the example that has multiple citations in a given context so that we can present the qualitative analysis well by investigating the top-$10$ ranked predictions. As shown in Table \ref{tab:qualitative analysis}, we consider an excerpt from \citet{liu2020roberta} that contains five citations. As we can see that \texttt{Symtax} correctly recommend three citations in the top-10, whereas \texttt{HAtten} only recommend one citation correctly at rank $1$ and \texttt{BM25} only suggest one correct citation at rank $10$. The use of title is crucial to performance, as we can see that many recommendations consist of the words ``BERT" and ``Pretraining", which are the keywords present in the title. One more observation is that the taxonomy plays a vital role in recommendations. The taxonomy category of the query is `Computation and Language`, and most of the recommended articles are from the same category. \texttt{SymTax} gives only one recommendation (Deep Residual Learning for Image Recognition) from a different category, i.e.``Computer Vision", whereas \texttt{HAtten} recommends three citations from different categories, i.e. (Deep Residual Learning for Image Recognition) from ``Computer Vision" and (Batch Normalization, and Adam) from ``Machine Learning".
\section{Conclusion}
In this paper, we present a model for local citation recommendation that leverages the notion of \textit{Symbiosis} from Biology, and we draw its analogy with human citation behaviour. We propose the notion of taxonomy fusion for learning rich concept representations and project them into hyperbolic space to derive a latent feature. We introduce a novel dataset that is comparatively large, dense, recent and more challenging than other existing datasets. Through several experiments and analyses, we prove our model as highly modular, which can run on datasets with comparatively few signals and accommodate additional signals as well. Our model consistently outperforms SOTA by huge margins for all evaluation metrics across all datasets.

\section{Limitations}
The current work marks the initial step towards incorporating human behaviour in designing a recommendation system for citation. We show empirically that such an inclusion leads to significant gains in performance. However, additional signals that resemble the actual citation behaviour can be incorporated to yield better performance. In the current setting, our system is limited to work in offline mode. We intend to transform our system to operate in the online setting, providing real-time recommendations.

\section{Ethics Statement}

Our work focuses on advancing citation recommendation and assisting the researchers in their academic writing process, where we are committed to maintain ethical standards. We will release our curated dataset and it can serve as a large and suitable benchmark for future research. Upholding transparency, our methodologies adhere to ethical guidelines, ensuring the responsible considerations. We assert that our work contributes positively to the citation ecosystem without raising ethical or moral concerns. We remain vigilant in addressing any unforeseen ethical challenges, driven by a commitment to principled research conduct. Our goal is to foster collaboration, uphold privacy, and enhance scholarly discourse.

\bibliography{cameraready}

\newpage

\appendix

\section{Appendix}
\label{sec:appendix}

We conduct another quantitative analysis using the section heading as an additional signal in our reranking module.

\subsection{Additional Experiment}
We concatenate the section heading with query context in reranker and run our two \texttt{SymTax} variants. From Table \ref{tab:section as a feature analysis}, we can observe that using section heading leads to a significant performance drop in SciBERT\_vector for all the metrics. However, for SPECTER\_graph, the overall performance remains nearly the same. Both of these patterns clearly indicate that using section heading as a feature acts as a noise, and thus the citation contexts are already rich. Since our proposed dataset contains this additional feature, it is suitable for two additional tasks: context-specific citation generation \citep{wang2022disencite}, and section heading prediction for a given citation context.

\subsection{Implementation Details}
We run all experiments on an NVIDIA DGX A100 GPU cluster, and our model is highly efficient in that it only requires about $5$GB of GPU memory for training \texttt{SymTax}. We use Adam optimizer with $\beta 1 = 0.9$ and $\beta 2 = 0.999$. We set learning rate to $1e^{ - 4}$ and weight decay to $1e^{ - 5}$ in prefetcher, whereas for reranker these were set to $1e^{ - 5}$ and to $1e^{ - 2}$ respectively for fine-tuning LM. We choose the top $100$ candidates returned by prefetcher as input to \texttt{Enricher}. We choose random seed $=12$ for sampling $10$k citation contexts from \texttt{ArSyTa} for conducting Quantitative Analysis as discussed in Table $4$ and Table $5$. We set margin $m=0.1$ in the triplet loss function. The maximum sequence length for LM is $512$. The values of $D$ and $d$ in the reranker are $768$ and $512$ respectively. Since \texttt{ArSyTa} is a highly dense network, we sort the enriched candidate list by frequency count and take the top $300$ candidates with the highest frequency count to run the reranker further. All the results are reported as an average of $3$ runs.

\subsection{Datasets}
\paragraph{ACL-200.}
This dataset contains papers published at ACL venues. It is a processed version of the ACL-ARC dataset created using ParsCit\footnote{https://github.com/knmnyn/ParsCit}, a string parsing package based on conditional random field. It contains citation contexts by considering $\pm 200$ characters around the citation placeholder.

\paragraph{FullTextPeerRead.}
It is an expansion of PeerRead dataset that contains the peer reviews of papers submitted to top venues in the Artificial Intelligence domain. So, \texttt{FullTextPeerRead} contains the citation contexts from the papers present in the PeerRead dataset.

\input{Tables/quant2}

\paragraph{RefSeer.}
This dataset is curated by extracting scientific articles belonging to various engineering domains. A citation excerpt is taken as the text of $\pm 200$ characters around the citation marker. It is a large dataset that contains 3.7 million citation contexts.

\paragraph{arXiv (HAtten).}
It is created using arXiv papers from a large and diverse corpus of scientific articles contained in S2ORC\footnote{https://github.com/allenai/s2orc}. For every paper having its full text available, a citation excerpt is considered if the cited paper is also present in the arXiv database. Following the similar trend setup by \texttt{ACL-200} and \texttt{RefSeer}, this dataset is also curated by considering the words in the $\pm 200$ character window around the citation marker.


\begin{figure}[htp!]
    \centering
    \includegraphics[width=\columnwidth]{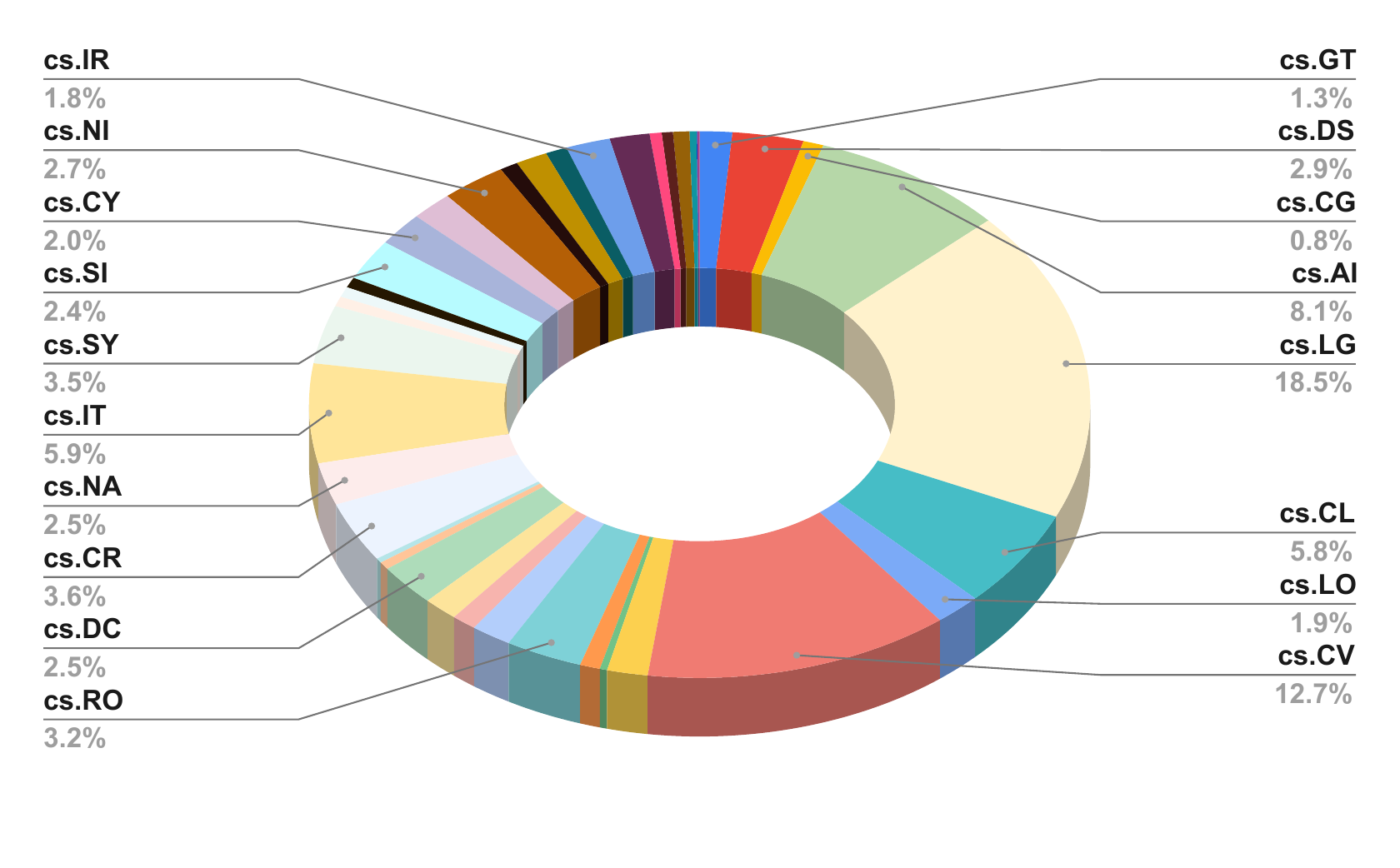}
    \caption{Statistics show the distribution of major category classes of flat-level arXiv taxonomy corresponding to \texttt{ArSyTa}. The highest number of research papers belong to Machine Learning (cs.LG), Computer Vision (cs.CV), and Artificial Intelligence (cs.AI). 
    }
    \label{fig:category_dist}
    \vspace{-4mm}
\end{figure}

\end{document}

%% file: Tables/datastats.tex
\begin{table}[!t]\centering
\renewcommand{\arraystretch}{1.5}
\resizebox{\columnwidth}{!}{
\begin{tabular}{l|cccc|ccccc}
\toprule

\multirow{2}{*}{Dataset} & \multicolumn{4}{c|}{\# Contexts}& \multirow{2}{*}{\# Papers}& \multirow{2}{*}{LCC}  & \multirow{2}{*}{Deg} &\multirow{2}{*}{Pub Years}  \\ 
\cmidrule{2-5}
 & Train & Val & Test& Total &  && & \\ 
\midrule
\textbf{ACL-200} & 30,390 & 9,381 & 9,585 & 49,356 & 19,776 & 0.035 & 3.42 & 2009-2015 \\ 
\textbf{FTPR} & 9,363 & 492 & 6,814 & 16,669 & 4,837 & 0.036 & 2.84 &2007-2017\\ 
\textbf{RefSeer} & 3,521,582 & 124,911 & 126,593 & 3,773,086 & 624,957 & 0.033 & 4.46 &-2014 \\ 
\textbf{arXiv} & 2,988,030 & 112,779 & 104,401 & 3,205,210 & 1,661,201 & 0.027 & 3.30 &1991-2020  \\ 
\midrule
\textbf{ArSyTa} & 8,030,837 & 124,188 & 124,189 & {\bf 8,279,214} & 474,341 & {\bf 0.051} & {\bf 9.98} & {\bf 2007-2023}  \\ 
\bottomrule
\end{tabular}}
\caption{Statistics across various datasets indicate the largest, densest and most recent nature of our dataset, \texttt{ArSyTa}. FTPR is \texttt{FullTextPeerRead}, arXiv is \texttt{arXiv(HAtten)}, and LCC and Deg are the average local clustering coefficient and average degree of the citation context network, respectively.}
\label{tab:data-stats}
\vspace{-3mm}
\end{table}

%% file: Tables/mainRes.tex
\begin{table}[!t]\centering
\resizebox{\columnwidth}{!}{
\begin{tabular}{l|llllll}
\toprule
\textbf{Model}                                                     & \textbf{R@5} & \textbf{R@10} & \textbf{R@20} & \textbf{R@50} & \textbf{NDCG} & \textbf{MRR} \\ 
\midrule
\multicolumn{7}{c}{\texttt{ACL-200}}                                                                                                                                      \\ 
\midrule
\textbf{\texttt{BM25}}                                                               & 0.1374       & 0.1939        & 0.2531        & 0.3486        & 0.0808        & 0.1074       \\
\textbf{\texttt{SciNCL}}                                                               & 0.1517       & 0.2250        & 0.3176        & 0.4467        & 0.1044        & 0.0669       \\
\textbf{\texttt{HAtten}}                                                             & 0.4186       & 0.4997        & 0.5579        & 0.5962        & 0.3002        & 0.2362       \\
\textbf{\texttt{SymTax} (SpecG)} & 0.4529       & 0.5897        & 0.7038        & 0.8396        & 0.3034        & 0.2126       \\
\textbf{\texttt{SymTax} (SciV)} & \textbf{0.5302}       & \textbf{0.6529}        & \textbf{0.7640}        & \textbf{0.8803}        & \textbf{0.3818}        & \textbf{0.2955}       \\ 
\midrule
\multicolumn{7}{c}{\texttt{FullTextPeerRead}}                                                                                                                             \\ 
\midrule
\textbf{\texttt{BM25}}                                                               & 0.2688       & 0.3371        & 0.4076        & 0.5168        & 0.1750        & 0.2136       \\
\textbf{\texttt{SciNCL}}
& 0.2173       & 0.3104        & 0.4217        & 0.5722        & 0.1452        & 0.0935       \\
\textbf{\texttt{HAtten}}                                                             & 0.5027       & 0.5788        & 0.6263        & 0.6514        & 0.3566        & 0.2847       \\
\textbf{\texttt{SymTax} (SpecG)}  & 0.4611       & 0.6266        & 0.7619        & 0.8899        & 0.3087        & 0.2090       \\
\textbf{\texttt{SymTax} (SciV)} & \textbf{0.6216}       & \textbf{0.7505}        & \textbf{0.8398}        & \textbf{0.9294}        & \textbf{0.4472}        & \textbf{0.3500}       \\ 
\midrule
\multicolumn{7}{c}{\texttt{RefSeer}}                                                                                                                                      \\ 
\midrule
\textbf{\texttt{BM25}}                                                               & 0.1737       & 0.2192        & 0.2677        & 0.3365        & 0.1185        & 0.1424       \\
\textbf{\texttt{SciNCL}}
& 0.0967       & 0.1486        & 0.2114        & 0.3089        & 0.0700        & 0.0450 \\
\textbf{\texttt{HAtten}}                                                             & 0.2672       & 0.3374        & 0.3985        & 0.4637        & 0.1925        & 0.1466       \\
\textbf{\texttt{SymTax} (SpecG)}  &  0.2724            &  0.3831             &   0.4960            &  0.6512             &  0.1942             &  0.1353            \\
\textbf{\texttt{SymTax} (SciV)}  & \textbf{0.3721}       & \textbf{0.4845}        & \textbf{0.5916}        & \textbf{0.7264}        & \textbf{0.2676}        & \textbf{0.1993}       \\ 
\midrule
\multicolumn{7}{c}{\texttt{arXiv(HAtten)}}                                                                                                                               \\ 
\midrule
\textbf{\texttt{BM25}}                                                               & 0.1529       & 0.1973        & 0.2455        & 0.3160        & 0.1019        & 0.1245       \\
\textbf{\texttt{SciNCL}} 
& 0.1076       & 0.1604        & 0.2227        & 0.3192        & 0.0737        & 0.0468 \\
\textbf{\texttt{HAtten}}                                                             & 0.2426       & 0.3292        & 0.4097        & 0.4949        & 0.1651        & 0.1136       \\
\textbf{\texttt{SymTax} (SpecG)}   & \textbf{0.2905}             &  \textbf{0.4095}             &  0.5308             &  \textbf{0.6992}             &  \textbf{0.1983}             &  \textbf{0.1323}            \\
\textbf{\texttt{SymTax} (SciV)}  & 0.2817      & 0.3997        & \textbf{0.5317}        & 0.6987        & 0.1928        & 0.1284       \\ 
\midrule
\multicolumn{7}{c}{\texttt{ArSyTa}}                                                                                                                                       \\ 
\midrule
\textbf{\texttt{BM25}}                                                               & 0.1777       & 0.2203        & 0.2640        & 0.3269        & 0.1155        & 0.1006       \\
\textbf{\texttt{SciNCL}} 
& 0.1612       & 0.2155        & 0.2757        & 0.3624        & 0.1088        & 0.0751 \\
\textbf{\texttt{HAtten}}                                                             & 0.1567       & 0.2046        & 0.2522        & 0.3070        & 0.1074        & 0.0766       \\
\textbf{{\texttt{SymTax}} (SpecG)}   &  0.2061 &  0.2747  &  0.3499  &  0.4668  &  0.1421  &  0.1003  \\
\textbf{\texttt{SymTax} (SciV)} & \textbf{0.2178} & \textbf{0.2976} & \textbf{0.3808}  & \textbf{0.5029}  & \textbf{0.1486} & \textbf{0.1018}  \\ 
\bottomrule
\end{tabular}}
\caption{Results clearly show that \texttt{SymTax} consistently outperforms SOTA (\texttt{HAtten}) across datasets on all metrics. Best results are highlighted in bold. Abbreviation: SpecG:- SPECTER\_Graph; SciV:- SciBERT\_Vector; R:- Recall.}
\label{tab:main results}
\end{table}

%% file: Tables/ablation.tex
\begin{table}[!t]\centering
\renewcommand{\arraystretch}{1.2}
\resizebox{\columnwidth}{!}{
\begin{tabular}{l|cccccc}
\toprule
\textbf{\texttt{SymTax} Variant}                                                            & \textbf{R@5} & \textbf{R@10} & \textbf{R@20} & \textbf{R@50} & \textbf{NDCG} & \textbf{MRR} \\ 
\toprule
\textbf{SciBERT\_vector}                                                           & 0.2178       & 0.2976        & 0.3808        & 0.5029        & 0.1486        & 0.1018       \\
\qquad {\bf -- Symbiosis} & 0.1794       & 0.2234        & 0.2651        & 0.3105        & 0.1194        & 0.0860        \\
\qquad {\bf -- Taxonomy}  & 0.1614       & 0.2377        & 0.3215        & 0.4611        & 0.1162        & 0.0787       \\
\qquad {\bf -- Hyperbolic}  & 0.1905       & 0.2678       & 0.3507       & 0.4719       & 0.1316       & 0.0891     \\
\midrule
\textbf{SPECTER\_graph}                                                            & 0.2061       & 0.2747        & 0.3499        & 0.4668        & 0.1421        & 0.1003       \\
\qquad {\bf -- Symbiosis}  & 0.1749       & 0.2178        & 0.2598        & 0.3079        & 0.1181        & 0.0862       \\
\qquad {\bf -- Taxonomy}   & 0.1795 & 0.2507 & 0.3384 & 0.4733 & 0.1263 & 0.0874   \\ 
\qquad {\bf -- Hyperbolic}  & 0.2028      & 0.2669       & 0.3444        & 0.4641       & 0.1386       & 0.0981      \\ 
\bottomrule 
\end{tabular}}
\caption{Ablation shows importance of \textit{Symbiosis}, taxonomy fusion and hyperbolic space on \texttt{ArSyTa}. Excluding Symbiosis reduces the metrics more as compared to the exclusion of taxonomy and hyperbolic space.}
\label{tab:ablation study}
\end{table}

%% file: Tables/quant1.tex
\begin{table}[!t]\centering
\renewcommand{\arraystretch}{1.2}
\resizebox{\columnwidth}{!}{
\begin{tabular}{l|cccccc}
\toprule
\textbf{\texttt{SymTax} Variant}  & \textbf{R@5} & \textbf{R@10} & \textbf{R@20} & \textbf{R@50} & \textbf{NDCG} & \textbf{MRR} \\ 
\toprule
\textbf{SciBERT\_graph}  & 0.1589       & 0.2421        & 0.3310        & 0.4607        & 0.1116        & 0.0715       \\
\textbf{SPECTER\_vector} & 0.1552       & 0.2324        & 0.3252        & 0.4607        & 0.1092        & 0.0712       \\
\textbf{SPECTER\_graph}  & \textit{0.2025}       & \textit{0.2667}        & \textit{0.3456}        & \textit{0.4654}        & \textit{0.1385}        & \textbf{0.0981}    \\
\midrule
\textbf{SciBERT\_vector} & \textbf{0.2097}       & \textbf{0.2876}        & \textbf{0.3754}        & \textbf{0.4939}        & \textbf{0.1432}        & \textit{0.0978}       \\
\bottomrule
\end{tabular}}
\caption{Analysis on choice of LM and taxonomy fusion on 10k random samples from \texttt{ArSyTa}. Best results are highlighted in bold and second best are italicised.}
\label{tab:LM and fusion choice analysis}
\end{table}

%% file: Tables/qual.tex
\begin{table*}[!t]\centering
\parbox{\dimexpr\textwidth-2\fboxsep}{\footnotesize{{\bf Citation Context:-} ``Self-training methods such as ELMo (Peters et al., 2018), GPT (Radford et al., 2018), BERT (Devlin et al., 2019), XLM (Lample and Conneau, 2019), and XLNet (Yang et al., 2019) have brought significant performance gains, but it can be challenging to determine which aspects of the methods contribute the most. Training is computationally expensive, limiting the amount of tuning that can be done, and is often done with private training data of varying sizes, limiting our ability to measure the effects of the modeling advances."} \\
{\bf Query Title:-} RoBERTa: A Robustly Optimized BERT Pretraining Approach}
\resizebox{\textwidth}{!}{
\begin{tabular}{lp{18em}p{18em}p{18em}p{18em}}
\toprule
\textbf{{\bf\#}} & \multicolumn{1}{c}{\textbf{BM25 recommendation}} & \multicolumn{1}{c}{\textbf{HAtten recommendation}} & \multicolumn{1}{c}{\textbf{SymTax recommendation}}\\ 
\midrule
1 & Sentence Encoders on STILTs: Supplementary Training on Intermediate Labeled-data Tasks & \textbf{BERT: Pre-training of Deep Bidirectional Transformers for Language Understanding} & \textbf{BERT: Pre-training of Deep Bidirectional Transformers for Language Understanding}  \\ \midrule
2             & Language-agnostic BERT Sentence Embedding                                                          & Deep Residual Learning for Image Recognition                                                                      & BART: Denoising Sequence-to-Sequence Pre-training for Natural Language Generation, Translation, and Comprehension \\ \midrule
3             & FlauBERT: Unsupervised Language Model Pre-training for French                                      & BART: Denoising Sequence-to-Sequence Pre-training for Natural Language Generation, Translation, and Comprehension & Deep Residual Learning for Image Recognition                                                                      \\ \midrule
4             & Passage Re-ranking with BERT                                                                       & Batch Normalization: Accelerating Deep Network Training by Reducing Internal Covariate Shift                      & ALBERT: A Lite BERT for Self-supervised Learning of Language Representations                                      \\ \midrule
5             & Does BERT Make Any Sense? Interpretable Word Sense Disambiguation with Contextualized Embeddings   & Neural Machine Translation by Jointly Learning to Align and Translate                                             & SuperGLUE: A Stickier Benchmark for General-Purpose Language Understanding Systems                                \\ \midrule
6             & BERTweet: A pre-trained language model for English Tweets                                          & Adam: A Method for Stochastic Optimization                                                                        & StructBERT: Incorporating Language Structures into Pre-training for Deep Language Understanding                   \\ \midrule
7             & Can You Tell Me How to Get Past Sesame Street? Sentence-Level Pretraining Beyond Language Modeling & Unified Language Model Pre-training for Natural Language Understanding and Generation                             & MPNet: Masked and Permuted Pre-training for Language Understanding                                                \\ \midrule
8             & Pre-trained language models as knowledge bases for Automotive Complaint Analysis                   & SuperGLUE: A Stickier Benchmark for General-Purpose Language Understanding Systems                                & \textbf{Cross-lingual Language Model Pretraining}                                                                 \\ \midrule
9             & Cross-Lingual BERT Transformation for Zero-Shot Dependency Parsing                                 & BERT and PALs: Projected Attention Layers for Efficient Adaptation in Multi-Task Learning                         & \textbf{XLNet: Generalized Autoregressive Pretraining for Language Understanding}                                 \\ \midrule
10            & \textbf{Cross-lingual Language Model Pretraining}                                                  & StructBERT: Incorporating Language Structures into Pre-training for Deep Language Understanding                   & Cross-Lingual BERT Transformation for Zero-Shot Dependency Parsing \\                                         
\bottomrule
\end{tabular}}
\caption{The table shows the top-10 citation recommendations given by various algorithms for a randomly chosen example from \texttt{ArSyTa}. Valid predictions are highlighted in bold. It clearly shows that \texttt{SymTax} (SciBERT\_vector) is able to recommend three valid articles in the top-10. In contrast, each of the \texttt{HAtten} and \texttt{BM25} could recommend only one valid article for the given citation context. \# denotes the rank of the recommended citations.}
\label{tab:qualitative analysis}
\end{table*}

%% file: Tables/quant2.tex
\begin{table}[!t]\centering
\renewcommand{\arraystretch}{1.2}
\resizebox{\columnwidth}{!}{
\begin{tabular}{l|cccccc}
\toprule
\textbf{\texttt{SymTax} Variant}                                                           & \textbf{R@5} & \textbf{R@10} & \textbf{R@20} & \textbf{R@50} & \textbf{NDCG} & \textbf{MRR} \\ \toprule
\textbf{SciBERT\_vector}                                                          & 0.2097       & 0.2876        & 0.3754        & 0.4939        & 0.1432        & 0.0978       \\
\qquad {\bf + Section} & 0.1556       & 0.2289        & 0.3193        & 0.4707        & 0.1123        & 0.0763       \\
\midrule
\textbf{SPECTER\_graph}                                                           & 0.2025       & 0.2667        & 0.3456        & 0.4654        & 0.1386        & 0.0981       \\
\qquad {\bf + Section}  & 0.2005       & 0.2772        & 0.3632        & 0.5001        & 0.1385        & 0.0950       \\ 
\bottomrule
\end{tabular}}
\caption{Analysis on the inclusion of section heading as a feature on 10k random samples from  \texttt{ArSyTa} data. The results indicate that using section heading as a feature acts as a noise as the citation contexts are already rich.}
\label{tab:section as a feature analysis}
\end{table}